\documentclass{ws-mpla}
\usepackage[super]{cite}
\usepackage{url}
\usepackage{graphicx}
\usepackage{epsfig}
\usepackage{epstopdf}
\usepackage{epsf}
\usepackage{amssymb}
\usepackage{tensor}
\usepackage{yhmath}
\usepackage{color}
\usepackage{float}

\begin{document}

\markboth{C. M. Nieto \& Y. Rodr\'{\i}guez}
{Massive Gauge-flation}

\catchline{}{}{}{}{}

\title{MASSIVE GAUGE-FLATION}

\author{\footnotesize CARLOS M. NIETO}

\address{Theoretical Particle Physics Group, SISSA (Scuola Internazionale Superiore di Studi Avanzati),\\
Via Bonomea 265, I-34136 Trieste, Italy\\
and\\
High Energy, Cosmology, and Astroparticle Physics Group, ICTP (The Abdus Salam International Centre for Theoretical Physics),\\ 
Strada Costiera 11, I-34151, Trieste, Italy\\ 
cnieto@sissa.it}

\author{YEINZON RODR\'IGUEZ}

\address{Centro de Investigaciones en Ciencias B\'asicas y Aplicadas, Universidad Antonio Nari\~no,\\ 
Cra 3 Este \# 47A-15, Bogot\'a D.C. 110231, Colombia\\
and\\
Escuela  de  F\'isica,  Universidad  Industrial  de  Santander,\\ 
Ciudad  Universitaria,  Bucaramanga  680002,  Colombia\\
and\\
Simons Associate at ICTP 
(The Abdus Salam International Centre for Theoretical Physics),\\ 
Strada Costiera 11, I-34151, Trieste, Italy\\ 
yeinzon.rodriguez@uan.edu.co}

\maketitle

\pub{Received 16 February 2016}{Revised 31 May 2016}


\begin{abstract}
Gauge-flation model at zeroth order in cosmological perturbation theory offers an
interesting scenario for realizing inflation within a particle physics context,
allowing us to investigate interesting possible connections between inflation and 
the subsequent evolution of the Universe. Difficulties, however, arise at the perturbative level,
thus motivating a modification of the original model. 
In order to agree with the latest Planck observations,
we modify the model such that the new dynamics can produce a relation between the spectral 
index $n_{s}$ and the tensor-to-scalar ratio $r$ allowed by the data. By 
including an identical mass term for each of the fields of the system, we find interesting 
dynamics leading to slow-roll inflation of the right length.
The presence of the mass term has the potential to modify
the $n_{s}$ vs. $r$ relation so as to agree with the data. 
As a first step, we study the model at zeroth order in 
cosmological perturbation theory, finding the conditions required for slow-roll inflation 
and the number of e-foldings of inflation.
Numerical solutions are used to explore the impact of the mass term. 
We conclude that the massive version of Gauge-flation offers a viable inflationary model.

\ccode{Preprint PI/UAN-2016-592FT}

\keywords{Non-Abelian gauge fields; inflation.}
\end{abstract}

\ccode{PACS Nos.: 11.15.-q; 98.80.Cq.}

\section{Introduction}	

The anomalies in the cosmic microwave background (CMB) map reported by Planck at low statistical 
significance\cite{Adeiso2,Adeiso} suggest modifications to the simplest scalar field inflation models since they 
cannot explain, for instance, the possible preferred direction in the 
Universe\cite{Adeiso,Groeneboomb,kim,ramazanov}.  Among the different anomalies involving a preferred direction, the dipolar power asymmetry in the CMB sky on large angular scales has the largest statistical significance (around $3\sigma$) with an amplitude of about 0.07.\cite{Adeiso2,Adeiso}  These anomalies could be explained by field theories 
involving vector or gauge fields, which make it possible to construct inflationary models that generate 
statistical anisotropy\cite{dimopoulos,golovnev,Malekreview}. Among the proposed models using vector 
fields, the proposal based on a gauge invariant theory is particularly appealing. This model, called 
\emph{Gauge-flation}\cite{Maleknejadd,Maleknejad,Malekreview} consists of an $SU(2)$ invariant 
Lagrangian with a Yang-Mills term as well as a contribution from the one-loop effective action called 
the $\kappa$-term. The latter term modifying the Yang-Mills action is crucial to generating the 
accelerated expansion. The $SU(2)$ invariance introduces three gauge fields, which can be arranged, 
naturally, in a way that produces the spatial homogeneity and isotropy (via the homomorphism between the $SU(2)$ and 
the $SO(3)$ groups).\cite{Maleknejadd} Although the configuration may appear finely tuned, a dynamical 
system study performed by a collaboration involving the same authors\cite{Maleksoda} demonstrated that a 
simple anisotropic initial configuration quickly evolves toward an isotropic configuration. This model 
exhibits interesting dynamics resulting from combining the Yang-Mills term and the $\kappa$-term.  The 
solution to the field equations shows that the combined behaviour of these terms generates slow-roll 
inflation. Moreover, this inflationary epoch lasts long enough to solve the standard cosmological 
classic problems.\footnote{Other very interesting models that make use of vector fields in gauge field theories, but this time with spontaneous symmetry breaking, are those presented in refs. \refcite{rinaldi1} and \refcite{rinaldi2}.  Each of these models is able to reproduce the late-time behaviour of the Universe as well as the primordial inflationary period or the other stages of its thermal history.}

Using cosmological perturbation theory, these authors found the range of parameters of the model and 
initial conditions for which the values of the spectral index $n_{s}$ and tensor-to-scalar ratio $r$ 
agree with the 7-year WMAP results.\cite{wmap} However, later studies showed that, according to the 
Planck 2013 results\cite{Oldplanck}, the conclusions in ref.~\refcite{Maleknejadd} are not correct since 
no value of $\gamma$ (a quantity relating the magnitude of the gauge fields, the gauge coupling $g$, 
and the Hubble parameter $H$) gives $r$ and $n_{s}$ in the allowed region for these parameters.\cite{Peloso}
Moreover, a tachyonic instability not reported in the original papers was found for the region 
$\gamma<2.$  More concretely, the model is ruled out in the stable region $\gamma \geq 2$ since either $n_s$ is too low (when $\gamma$ is small) or $r$ is too high (when $\gamma$ is large).

In order to retain the interesting properties of the Gauge-flation model at the background level, we 
modify the original proposal to alter the predictions for $n_{s}$ and $r$ as well to avoid the tachyonic 
instability. By introducing a mass term in the Lagrangian, we induce a longitudinal mode in the field 
perturbations which can be expected to modify the allowed region for $r$ and $n_{s}$. This modification 
is beneficial since the spectral index acquires a new contribution from the slow-roll parameter $\eta$ 
(i.e., from the mass term).  This paper studies the new model at the background level whereas the study of the 
perturbations is deferred to a subsequent publication. We want to know whether including the mass term 
can give the right amount of inflationary expansion and to determine the conditions required for 
accelerated expansion in the new model.

There exists concern regarding quantum and classical instabilities generated in massive vector field 
models of inflation\cite{himmetoglu}. However, refs.~\refcite{dulaney} and \refcite{carroll} show that a 
Maxwell kinetic term for an Abelian gauge field is one of the three cases that avoid such instabilities. 
(See for example the successful vector curvaton scenario\cite{vcurvaton}).  In Gauge-flation, we are not 
dealing with an Abelian vector field, but we expect that an analysis similar to that carried out in 
refs. \refcite{dulaney} and \refcite{carroll} will demonstrate the absence of such instabilities.  
It is worth remarking that models with Maxwell-type kinetic term, a mass term, and a $\kappa$ term lead to second-order field 
equations, which is a necessary condition for avoiding the famous Ostrogradski 
instability\cite{ostrogradski} in the Abelian\cite{tasinato,heisenberg,allys,heisenberg2,allys3} and non-Abelian 
cases\cite{tasinato2,allys2}.

This paper is organized as follows.  Section \ref{sec:model} describes the model setup highlighting 
differences with the original model. In Section \ref{sec:motion}, we derive the gauge field and Einstein 
equations in order to establish the conditions for accelerated expansion. 
We also discuss gauge 
fixing. In Section \ref{sec:slow}, we obtain the conditions for slow-roll inflation and some important 
relations such as the amount of inflation. In Section \ref{sec:num}, we explore numerical solutions to 
the field and Einstein equations. We show that it is possible to get accelerated expansion with the 
desired properties and analyse the impact of the mass term in the model. Section \ref{sec:con} presents 
some conclusions and indicates directions for future work.

\section{Modified Gauge-flation}
\label{sec:model}

Gauge-flation was proposed as a model of inflation based on an $SU(2)$ symmetry 
and presents interesting 
features in the unperturbed regime \cite{Maleknejadd,Maleknejad,Malekreview}. 
For several initial conditions Gauge-flation produces a sufficient amount of inflation as
well as generating homogeneity and isotropy using an interesting field configuration.
Modifications to gravity are not needed.
Because of the important role that the non-Abelian gauge field theories play at describing the known fundamental interactions, it is clear that Gauge-flation represents a step ahead in the aim of merging inflationary cosmology and particle physics.  Indeed, the presence of non-Abelian gauge fields allows us to easily and adequately connect inflation with the physics of the subsequent evolution of the Universe 
(e.g., preheating\cite{deskins} and reheating\cite{ghalee}).
The model, however, exhibits some negative features at the perturbative level \cite{Peloso}. 
More concretely, no set of initial conditions generates combinations 
of $n_{s}$ and $r$ compatible with the Planck results. We propose here 
a modification of the model by including a mass term that might arise
from some Higgs-like mechanism in a more sophisticated setup (see, for example, ref. \refcite{tasinato2}).  We expect
this mass term to introduce a new longitudinal mode to modify the predictions for
$n_{s}$ and $r$. The theoretical prediction for $n_s$ would move to the 
region allowed by observations as the slow-roll parameter $\eta$, which contributes to $n_s$, gets a 
contribution from the mass term. In this paper, we only 
analyse the impact of the mass term on the inflationary dynamics. We also check whether it remains 
possible to produce a successful inflation and under what conditions.

The Lagrangian describing the system is given by
\begin{equation}\label{Newlag}
\mathcal{L}=-\frac{1}{4}F^{a}_{\mu\nu}F_{a}^{\mu\nu}+\frac{\kappa}{384}
(\varepsilon^{\mu\nu\lambda\sigma}F^a_{\mu\nu}F^{a}_{\lambda\sigma})^{2}
-\frac{1}{2}m^{2}A^{a}_{\ \mu}A_{a}^{\ \mu} \,,
\end{equation}
where $\varepsilon^{\mu\nu\lambda\sigma}$ is the totally antisymmetric Levi-Civita tensor and 
$\kappa/384$ is a positive dimensionful parameter related to the cutoff of the theory. 
The mass is the same for all three fields, the only choice which is consistent with the needed symmetries:
homogeneity and isotropy. On the other hand, this Lagrangian breaks the gauge symmetry explicitly,
and consequently we cannot apply a gauge fixing condition as was done for the original model. 
Nevertheless, we find a way to use the same field configuration as in the original massless 
Gauge-flation model as shown in the following Section. Using the flat Friedmann-Robertson-Walker (FRW) 
metric for the spacetime
\begin{equation}
 ds^{2}=-dt^{2}+a^{2}(t)\delta_{ij}dx^{i}dx^{j} \,,
\end{equation}
we use the orthonormal tetrad 
\begin{eqnarray}
 \hat{e}^{(0)}_{\ \ \mu}=(1,0,0,0) \,, && \ \  \hat{e}^{(1)}_{\ \ \mu}=(0,a(t),0,0) \,, \nonumber \\
 \hat{e}^{(2)}_{\ \ \mu}=(0,0,a(t),0) \,, &&  \ \  \hat{e}^{(3)}_{\ \ \mu}=(0,0,0,a(t)) \,,
\end{eqnarray}
to write the spatial components of the fields in the form $A^{a}_{\ i}=\hat{e}^{(j)}_{\ \ i}A^{a}_{\ 
(j)}$, where $A^{a}_{\ (j)}$ are the components of each gauge field $A^{a}_{\ i}$ along the orthonormal 
basis $\hat{e}^{(j)}_{\ \ i}$. We now use as an Ansatz a field configuration with the vectors 
$A^{a}_{\ (j)}$ oriented along each basis vector of the tetrad, in other words 
\begin{equation}\label{Ansatz}
    A^{a}_{\ (j)}=\psi(t)\delta^{a}_{\ (j)} \,.
\end{equation}
This implies that the three spatial vectors $A^{a}_{\ i}$ are orthonormal, namely 
$A^{a}_{\ i}=a(t)\psi(t)\delta^{a}_{\ i}=\phi(t)\delta^{a}_{\ i}$. As explained in 
ref.~\refcite{Maleksoda}, this Ansatz is an attractor solution of an anisotropic inflationary model described 
by the Bianchi type I metric. 
This configuration is, therefore, not a random choice.
We expect the same to happen in the case of massive fields since, as we will see 
later, it is possible to get inflation with similar properties to the original model. With the Ansatz in 
eq. (\ref{Ansatz}), we see that the system contains four variables coming from the three fields: 
$\psi(t)$ and the three temporal components of the fields $A^{a}_{\ 0}$. 
In the following Section, we 
study the dynamics of the system where we see, in particular, how to get a constraint equation for $A^{a}_{\ 0}$.

\section{Field and Einstein Equations}
\label{sec:motion}

We obtain the field and Einstein equations in order to study the inflationary dynamics.
As we will see in the following, the field equation for $A^{a}_{\ \mu}$ imposes an important constraint on the field configuration. 
The Euler-Lagrange equations
\begin{equation}\label{NewEL}
\frac{\partial(\sqrt{-\tilde{g}}\mathcal{L})}{\partial A_{\ \mu}^{a}}-\partial_{\nu}\left[\frac{\partial(\sqrt{-\tilde{g}}\mathcal{L})}{\partial(\partial_{\nu}A_{\ \mu}^{a})}\right]=0 \,,
\end{equation}
where $\tilde{g}$ is the determinant of the metric tensor $g_{\mu\nu}$ with 
the Lagrangian in eq. (\ref{Newlag}), can be rewritten as 
\begin{equation}
   D_{\mu}\frac{\partial\sqrt{-\tilde{g}}\mathcal{L}}{\partial F_{\ \mu\nu}^{a}}+m^{2}\sqrt{-\tilde{g}}A^{a}_{\ \mu}g^{\mu\nu}=0 \,,
    \label{Eulernew}
\end{equation}
where $D_{\mu}$ is the $SU(2)$ covariant derivative. For $\nu=0$, we obtain the equation
\begin{align}
   & \partial_{i}\left[F^{0i}_{\ a}-\frac{\kappa}{48}\varepsilon^{0i\lambda\sigma}F^{a}_{\ \lambda\sigma}(\varepsilon^{\rho\omega\gamma\kappa}F^{b}_{\ \rho\omega}F^{b}_{\ \gamma\kappa})\right]-m^{2}A^{a}_{\ 0} \\ \notag
   & -g\epsilon^{a}_{\ dc}A^{d}_{\ i}\left[F^{0i}_{\ c}-\frac{\kappa}{48}\varepsilon^{0i\lambda\sigma}F^{c}_{\ \lambda\sigma}(\varepsilon^{\rho\omega\gamma\kappa}F^{b}_{\ \rho\omega}F^{b}_{\ \gamma\kappa})\right]=0 \,,
\end{align}
where $\epsilon^{a}_{\ dc}$ 
(i.e., the totally antisymmetric three-dimensional Levi-Civita tensor)
represents the structure constants of the $SU(2)$ group and $g$ is the respective coupling constant. 
The homogeneity property of 
the background sets all the derivatives $\partial_{i}A^{a}_{\ \mu}$ to zero,
giving a constraint on the temporal components of the fields
\begin{equation}
    A^{a}_{\ 0}\left(m^{2}+\frac{2g^{2}\phi^{2}}{a^{2}}\right)=0 \,.
    \label{Tresult}
\end{equation}
This implies that all the zero components of the fields vanish,
and we thus recover the initial configuration 
of the fields given in ref.~\refcite{Maleknejadd}:
\begin{equation}
    A^{a}_{\ 0}=0 \,, \ \ A^{a}_{\ i}=\phi(t)\delta^{a}_{\ i} \,.
    \label{Newconf}
\end{equation}
We have just seen, therefore, that the explicit symmetry breaking of the SU(2) invariance does not alter the configuration in 
eq.~(\ref{Newconf}) provided that we assume homogeneity and isotropy. 
Consequently, the spatial components 
of the Euler-Lagrange equations reduce to
\begin{equation}\label{Gaugemotion}
    \left(1 + \kappa\frac{g^{4}\phi^{4}}{a^{4}}\right)\frac{\ddot{\phi}}{a}+\left(1 + \kappa\frac{\dot{\phi}^{2}}{a^{2}}\right)\frac{2g^{2}\phi^{3}}{a^{3}}+\left(1-3\kappa\frac{g^{2}\phi^{4}}{a^{4}}\right)\frac{H\dot{\phi}}{a}+m^{2}\frac{\phi}{a}=0 \,,
\end{equation}
where the dot indicates a derivative with respect to cosmic time.

On the other hand, the energy-momentum tensor 
\begin{equation}
T_{\mu\nu}=-2\frac{\delta\mathcal{L}}{\delta g^{\mu\nu}} + g_{\mu\nu}\mathcal{L} \,,
\end{equation}
for our model is
\begin{equation}
    T_{\mu\nu}=F^{a}_{\ \mu\beta}F^{a}_{\ \nu\alpha}g^{\beta\alpha}-\frac{\kappa}{192}\frac{g_{\mu\nu}}{(\sqrt{-\tilde{g}})^{2}}\left(\epsilon^{\alpha\beta\lambda\sigma}F^{a}_{\ \alpha\beta}F^{a}_{\ \lambda\sigma}\right)^{2}+m^{2}A^{a}_{\ \mu}A^{a}_{\ \nu}+g_{\mu\nu}\mathcal{L}_{red} \,,
\end{equation}
where $\mathcal{L}_{red}$ is the reduced Lagrangian of the model. This reduced Lagrangian is derived by
replacing the field configuration in eq.~(\ref{Newconf}) in the original Lagrangian in eq. 
(\ref{Newlag}):
\begin{equation}
    \mathcal{L}_{red}=\frac{3}{2}\left[\frac{\dot{\phi}^{2}}{a^{2}}-\frac{g^{2}\phi^{4}}{a^{4}}+ \kappa\frac{g^{2}\phi^{4}\dot{\phi}^{2}}{a^{6}}-\frac{m^{2}\phi^{2}}{a^{2}}\right] \,.
\end{equation}
The non-vanishing components of the energy-momentum tensor are
\begin{eqnarray}
    T_{00} &=& \frac{3}{2}\left[\frac{\dot{\phi}^{2}}{a^{2}}+\frac{g^{2}\phi^{4}}{a^{4}}+ \kappa\frac{g^{2}\phi^{4}\dot{\phi}^{2}}{a^{6}}+\frac{m^{2}\phi^{2}}{a^{2}}\right] \,, \\
    \label{Tzero}
    T_{ij} &=& \frac{1}{2}\left[\dot{\phi}^{2}+\frac{g^{2}\phi^{4}}{a^{2}} - 3\kappa\frac{g^{2}\phi^{4}\dot{\phi}^{2}}{a^{4}}-m^{2}\phi^{2}\right]\delta_{ij} \,.
    \label{Tij}
\end{eqnarray}
As expected, the Universe is represented by a perfect fluid with an energy-momentum tensor of the 
form $T_{\mu\nu}=(\rho+P)U_{\mu}U_{\nu}+Pg_{\mu\nu}$, $\rho$ being the energy density, $P$ the 
isotropic pressure, and $U_{\mu}$ the fluid 4-velocity. For a comoving observer, the 
expressions for the energy density and pressure are
\begin{eqnarray}
    \rho &=& \frac{3}{2}\left[\frac{\dot{\phi}^{2}}{a^{2}}+\frac{g^{2}\phi^{4}}{a^{4}}+ \kappa\frac{g^{2}\phi^{4}\dot{\phi}^{2}}{a^{6}}+\frac{m^{2}\phi^{2}}{a^{2}}\right] \,, \\
    P &=& \frac{1}{2}\left[\frac{\dot{\phi}^{2}}{a^{2}}+\frac{g^{2}\phi^{4}}{a^{4}} -3\kappa\frac{g^{2}\phi^{4}\dot{\phi}^{2}}{a^{6}}-\frac{m^{2}\phi^{2}}{a^{2}}\right] \,.
\end{eqnarray}
With these formulae, we obtain the Einstein equations for the model
\begin{eqnarray}
    H^{2} &=& \frac{1}{2m_{Pl}^{2}}\left[\frac{\dot{\phi}^{2}}{a^{2}}+\frac{g^{2}\phi^{4}}{a^{4}} + \kappa\frac{g^{2}\phi^{4}\dot{\phi}^{2}}{a^{6}}+\frac{m^{2}\phi^{2}}{a^{2}}\right] \,, \label{Einstein1} \\
    \dot{H} &=& - \frac{1}{m_{Pl}^{2}}\left[\frac{\dot{\phi}^{2}}{a^{2}}+\frac{g^{2}\phi^{4}}{a^{4}}+\frac{1}{2}\frac{m^{2}\phi^{2}}{a^{2}}\right] \,, \label{Einstein2}
\end{eqnarray}
where $m_{Pl}$ is the reduced Planck mass.

We now must solve eqs.~(\ref{Gaugemotion}), (\ref{Einstein1}), and (\ref{Einstein2}) to check 
whether the conditions for inflation are satisfied. However, we can simplify
the 
Einstein equations and gain more intuition by expressing the energy density as the sum 
of three contributions coming from the different terms in the Lagrangian in eq.~(\ref{Newlag}):
\begin{equation}\label{Newrho}
    \rho=\rho_{YM}+\rho_{\kappa}+\rho_{mass} \,,
\end{equation}
where 
\begin{equation}
\rho_{YM}=\frac{3}{2}\left[\frac{\dot{\phi}^{2}}{a^{2}}+\frac{g^{2}\phi^{4}}{a^{4}}\right] \,, \ \ \rho_{\kappa}=\frac{3}{2}\kappa\frac{g^{2}\phi^{4}\dot{\phi}^{2}}{a^{6}} \,, \ \ \rho_{mass}=\frac{3}{2}\frac{m^{2}\phi^{2}}{a^{2}} \,. \label{densities}
\end{equation}
The isotropic pressure takes the form
\begin{equation}\label{Newp}
    P=\frac{1}{3}\rho_{YM}-\rho_{\kappa}-\frac{1}{3}\rho_{mass} \,,
\end{equation}
and the condition for a period of accelerated expansion, given by $\rho+3P<0$, implies that
\begin{equation}\label{Tsamecon}
    \rho_{\kappa}>\rho_{YM}.
\end{equation}
The mass term {\em does not} affect the existence of an accelerated expansion,
which requires only that 
$\rho_{\kappa}>\rho_{YM}.$  
The inequality inverts as inflation comes to an end. 
An interesting dynamics results from 
the interplay of the contributions of the Yang-Mills term and the $\kappa$-term to the 
energy density. The latter starts inflation and the former ends
inflation. Although the 
mass term seems irrelevant to the overall behaviour during inflation, we will see in the following section that
it strongly influences its length.

\section{Slow-Roll Inflation}
\label{sec:slow}

We derive the conditions required for slow-roll inflation. As stated before, this type of inflation can easily lead to a sufficient length of 
accelerated expansion which is needed to solve the classic problems of the standard 
cosmology.\cite{uzan,Weinbergbook,LythII,Baumanna}. Moreover, slow-roll inflation 
guarantees a spectral index for the power spectrum of the primordial curvature 
perturbation very close to one, as needed to satisfy the observational restrictions 
presented in ref.~\refcite{Newplanck}.

The slow roll parameters
\begin{equation}\label{Setparameters}
\varepsilon\equiv-\frac{\dot{H}}{H^{2}},\ \ 
\eta\equiv-\frac{\ddot{H}}{2\dot{H}H},\ \ 
\delta\equiv-\frac{\dot{\psi}}{\psi H} \,,
\end{equation}
characterize the evolution 
of the Hubble parameter $H$ and the field $\psi=\phi/a$. 
It follows that
\begin{equation}
    \eta=\varepsilon-\frac{\dot{\varepsilon}}{2H\varepsilon} \,.
    \label{Etaeprel}
\end{equation}
Before giving the analytic expressions for $\varepsilon$, $\eta$ and $\psi$ in terms of the model
variables, we find 
the conditions on the energy density needed for slow-roll inflation, which
requires that $\varepsilon,|\eta|,\delta \ll 1$ 
\cite{uzan,LythII,Weinbergbook,Baumanna}. 
From eqs.~(\ref{Einstein1}), (\ref{Einstein2}), and (\ref{Setparameters}), we obtain
\begin{equation}
    \varepsilon=\frac{2\rho_{YM}+\rho_{mass}}{\rho_{\kappa}+\rho_{YM}+\rho_{mass}} \,.
    \label{Condslow}
\end{equation}
The condition on $\varepsilon$ for slow-roll inflation is satisfied if 
$\rho_{\kappa}\gg\rho_{YM}$ {\em and} $\rho_{\kappa}\gg\rho_{mass}$. We can see how
much 
the mass term affects slow-roll inflation: $\rho_{mass}$ must be much smaller than $\rho_{\kappa}$. 
Although $\rho_{mass}$ is not relevant for generating inflation, 
{\em it is} relevant if we want inflation to be slow-roll,
which in turn implies a significant 
influence of the mass on the length of inflation. 
Numerical simulations demonstrating
this fact will be presented in the following Section.

We now
study the conditions to ensure that $|\eta|,\delta \ll 1$, which is expressed as
conditions on $\varepsilon$ and its time derivative. 
From eq. (\ref{Etaeprel}), it follows that 
$|\eta|\sim\varepsilon\ll1$ only if $\dot{\varepsilon}/H\varepsilon\sim\varepsilon$, 
which we shall assume, i.e., if the relative variation of $\varepsilon$ is very small, then $|\eta|\ll1$. 
Using the Friedmann equation and eqs.~(\ref{densities}) and (\ref{Condslow}), 
we obtain
\begin{equation}
    \varepsilon=\frac{1}{H^{2}m_{Pl}^{2}}\left[\frac{\dot{\phi}^{2}}{a^{2}}+\frac{g^{2}\phi^{4}}{a^{4}}+\frac{m^{2}\phi^{2}}{2a^{2}}\right] \,,
\end{equation}
or equivalently
\begin{equation} 
    \varepsilon=\frac{\psi^{2}}{m_{Pl}^{2}}\left((1-\delta)^{2}+\gamma+\omega/2\right) \,,
    \label{Newep}
\end{equation}
where 
\begin{equation}
    \gamma\equiv\frac{g^{2}\psi^{2}}{H^{2}} \,, \ \ \omega\equiv\frac{m^{2}}{H^{2}} \,.
    \label{Twoparam}
\end{equation}
$\varepsilon$ also follows from the Friedmann equation and eqs.~(\ref{densities}) and 
(\ref{Condslow}), so that 
\begin{equation}\label{Usefulrel}
\varepsilon=2-\frac{\kappa g^{2}\psi^{6}
(1-\delta)^{2}}{m_{Pl}^{2}}-\frac{m^{2}\psi^{2}}{2H^{2}m_{Pl}^{2}} \,.
\end{equation}
If we derive the previous equation with respect to the cosmic time and using
eqs.~(\ref{Etaeprel}) and (\ref{Newep}), we find 
\begin{align} \label{Reletaepdel}
    \eta=\varepsilon-&\left[\frac{3\delta}{\varepsilon}+\frac{\dot{\delta}}{(1-\delta)H\varepsilon}\right]\left[2-\varepsilon-\frac{\varepsilon\omega/2}{((1-\delta)^{2}+\gamma+\omega/2)}\right] \\ \notag
    &+\frac{(\varepsilon-\delta)\omega/2}{((1-\delta)^{2}+\gamma+\omega/2)} \,. 
\end{align}
Therefore, from $|\eta| \sim\varepsilon$ and 
$\frac{\omega/2}{((1-\delta)^{2}+\gamma+\omega/2)}<1$, we conclude that 
$\delta\sim\varepsilon^{2}$ must be satisfied in order to have slow-roll inflation. Moreover, 
from this 
result and the condition on the time variation of $\varepsilon$ (i.e., $\dot{\varepsilon}\sim 
H\varepsilon^{2}$), we obtain that $\delta$ also varies slowly in time 
(i.e., $\dot{\delta}/H\delta\sim\varepsilon$).

Given the conditions for slow-roll inflation, we can now obtain expressions for the 
parameters $\varepsilon,\eta,\delta$ in the slow-roll approximation. Eq.~(\ref{Newep}) gives 
\begin{equation}
    \varepsilon\approx\frac{\psi^{2}}{m_{Pl}^{2}}(1+\gamma+\omega/2) \,.
    \label{Finalep}
\end{equation}
Combining this result with eqs.~(\ref{Etaeprel}) and (\ref{Twoparam}), we obtain 
\begin{equation}
    \eta\approx\frac{\psi^{2}}{m_{Pl}^{2}} \,.
    \label{Finaleta}
\end{equation}
Finally, using $({\omega/2})/{((1-\delta)^{2}+\gamma+\omega/2)}<1$, 
we rewrite eq.~(\ref{Reletaepdel}) as 
\begin{equation}
    \eta\approx\varepsilon-2\left[\frac{3\delta}{\varepsilon}+\frac{\dot{\delta}}{H\varepsilon}\right]+\frac{\varepsilon \omega/2}{(1+\gamma+\omega/2)} \,,
\end{equation}
which helps us to find an expression for $\delta$ using the previous results for $\varepsilon$ and $\eta$:
\begin{equation}
    \delta\approx\frac{(\gamma+\omega/2)}{6(1+\gamma+\omega/2)}\varepsilon^{2} \,.
    \label{Finaldel}
\end{equation}
These expressions for $\varepsilon$, $\eta$ and $\delta$ coincide with the behaviour described above.
For example, we see that $\delta\sim\varepsilon^{2}$.

One motivation for slow-roll is to produce enough $e$-folds
of inflation $N_e$. 
To calculate $N_{e}$, we take into account that $\delta=-\dot{\psi}/H\psi$ and 
$\delta\sim\varepsilon^{2}$, so that $\psi$ remains almost constant during the slow-roll inflationary 
period. This allows us to write, using eq.~(\ref{Finalep}),
\begin{equation}
    \frac{\varepsilon}{\varepsilon_{i}}\approx\frac{1 + \gamma+\omega/2}{1 + \gamma_{i}+\omega_i/2} \,,
\end{equation}
where the subscript $i$ denotes the beginning of slow-roll inflation. Since 
$\varepsilon=1$ at the end of inflation, 
\begin{equation}
    \gamma_f+\omega_f/2\approx\frac{1+\gamma_{i}+\omega_{i}/2}{\varepsilon_{i}} \,,  \label{Relepga}
\end{equation}
where the subscript $f$ denotes the end of slow-roll inflation. This expression is
useful to obtaining $N_{e}$.  The definition of $N_{e}$ between times $t_{i}$ 
and $t_{f}$ is
\begin{equation}
    N_{e}=\int_{t_{i}}^{t_{f}}Hdt \,,
\end{equation}
which using the change of variable $dt=dH/\dot{H}$ can be rewritten as
\begin{equation}
    N_{e}=-\int_{H_{i}}^{H_{f}}\frac{dH}{H\varepsilon} \,.
\end{equation}
To solve the integral, we introduce the variable $\theta=\gamma + \omega/2,$ finding that
\begin{equation}
    N_{e}\approx-\frac{m_{Pl}^2}{2\psi^{2}}
\int_{\theta_{i}}^{\theta_{f}}\frac{-1/\theta^{2}}{(1+\theta)/\theta}d\theta \,.
\end{equation}
Finally, we use eqs. (\ref{Finalep}) and (\ref{Relepga}) to obtain 
the following formula for 
$N_{e}$ in terms of the initial values of $\varepsilon$, $\gamma$ and $\omega$:
\begin{equation}
    N_{e}\approx\frac{1+\gamma_{i}+\omega_{i}/2}{2\varepsilon_{i}} \ln \left[\frac{1+\gamma_{i}+\omega_{i}/2}{\gamma_{i}+\omega_{i}/2}\right] \,.
    \label{Finalefolds}
\end{equation}
The final amount of inflation does not depend on the evolution of $\varepsilon$, $\gamma$, or 
$\omega ,$ but rather on their initial values.
From the initial conditions of the model, we can check whether or not 
the right amount of inflation is produced. Thus the initial conditions on $\psi$, 
$\dot{\psi}$, and $H$ are constrained such that the value of $N_{e}$ satisfies the required bound ($N_e 
\gtrsim 60$)\cite{uzan,Weinbergbook,LythII,Baumanna}.. In the next Section, we solve numerically the 
field and Einstein equations of the model and check the validity of eq. (\ref{Finalefolds}) as well 
as the impact of $\omega$ (the mass term) on $N_{e}$.

\section{Numerical Solution}
\label{sec:num}

We solve numerically the field and Einstein equations of the model, which
are eqs.~(\ref{Gaugemotion}), 
(\ref{Einstein1}), and (\ref{Einstein2}). Only two of these equations are independent. 
The idea is to check whether introducing the mass term allows the model to produce the necessary
accelerated expansion. By analysing the evolution of $\varepsilon$, we check whether the inequality in eq.~(\ref{Tsamecon}) is satisfied
and also whether the inflationary period is slow-roll.
Furthermore, we check the validity of eq.~(\ref{Finalefolds}) for the number of $e$-folds of inflation. 
The model has three parameters ($g$, $\kappa ,$ and $m$) and two dynamical degrees of freedom 
($\phi(t)$ and $a(t)$). We must therefore specify the initial values of $\phi$, $a$ 
and their first time derivatives.  As the physical field is $\psi=\phi/a$, and the measurable quantity is 
$H$ instead of $a$, we specify initial conditions in terms of $\psi$ and $H$. 
We then find the temporal evolution of $\psi$, $\varepsilon$, $\rho_\kappa/\rho_{YM}$, $\rho_\kappa/\rho_{mass}$, $\rho_{mass}/\rho_{YM}$, and $N$, as well as  $\dot{\phi}/a$ vs. $\psi$ for 
two sets of initial conditions. These initial conditions are chosen so that the expected 
properties of the model are obtained, keeping a similarity with the conditions chosen in the 
original massless Gauge-flation model\cite{Maleknejadd}.

\begin{figure}[h]
\centerline{\includegraphics[width=2.5in]{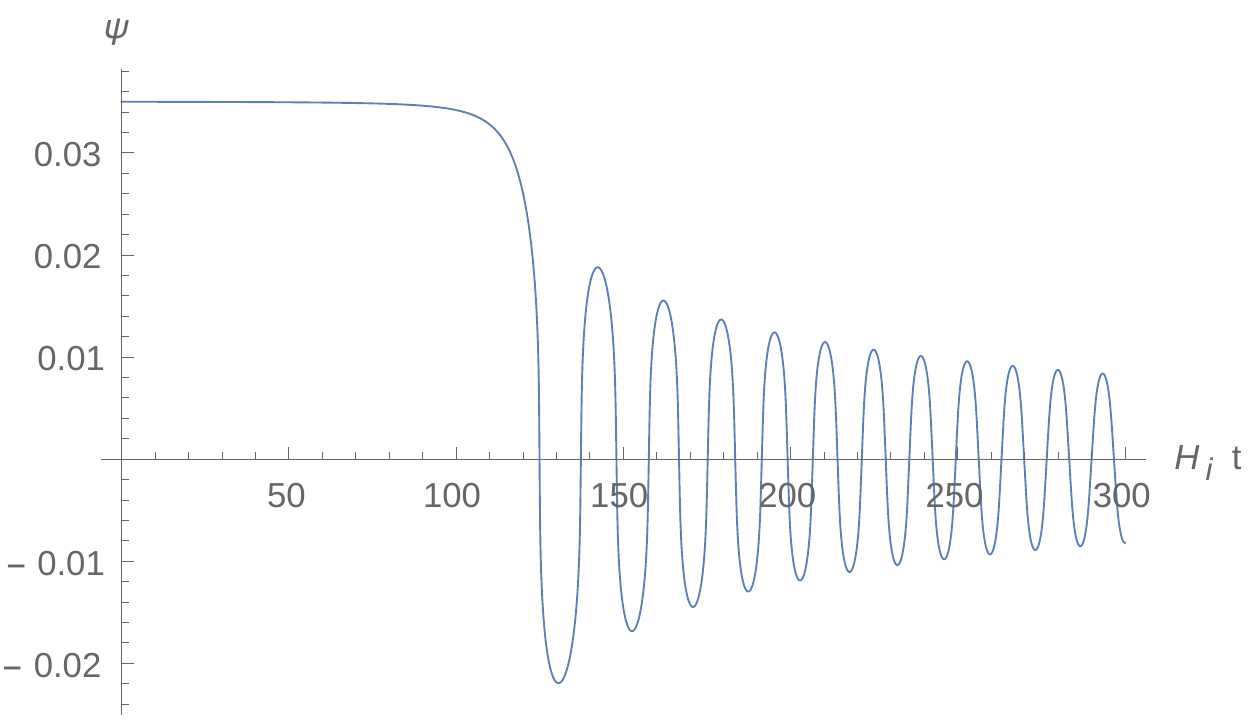}
\includegraphics[width=2.5in]{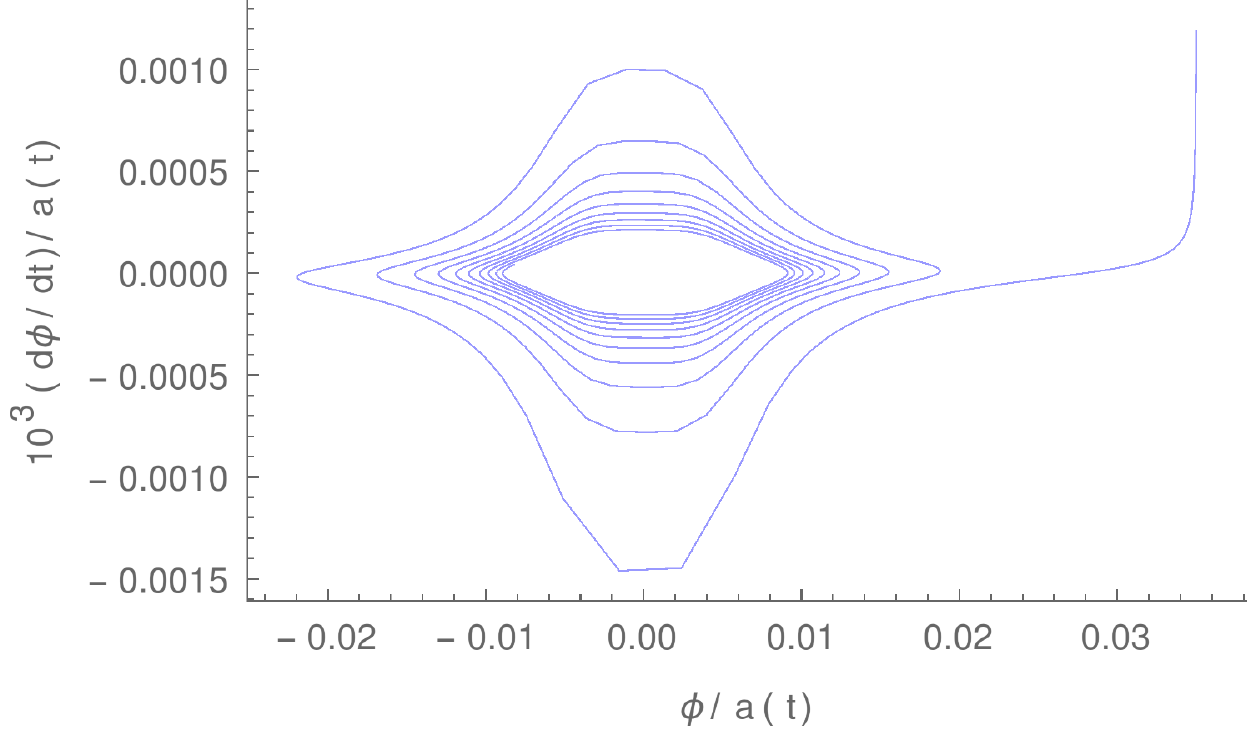}} 
\centerline{\includegraphics[width=2.5in]{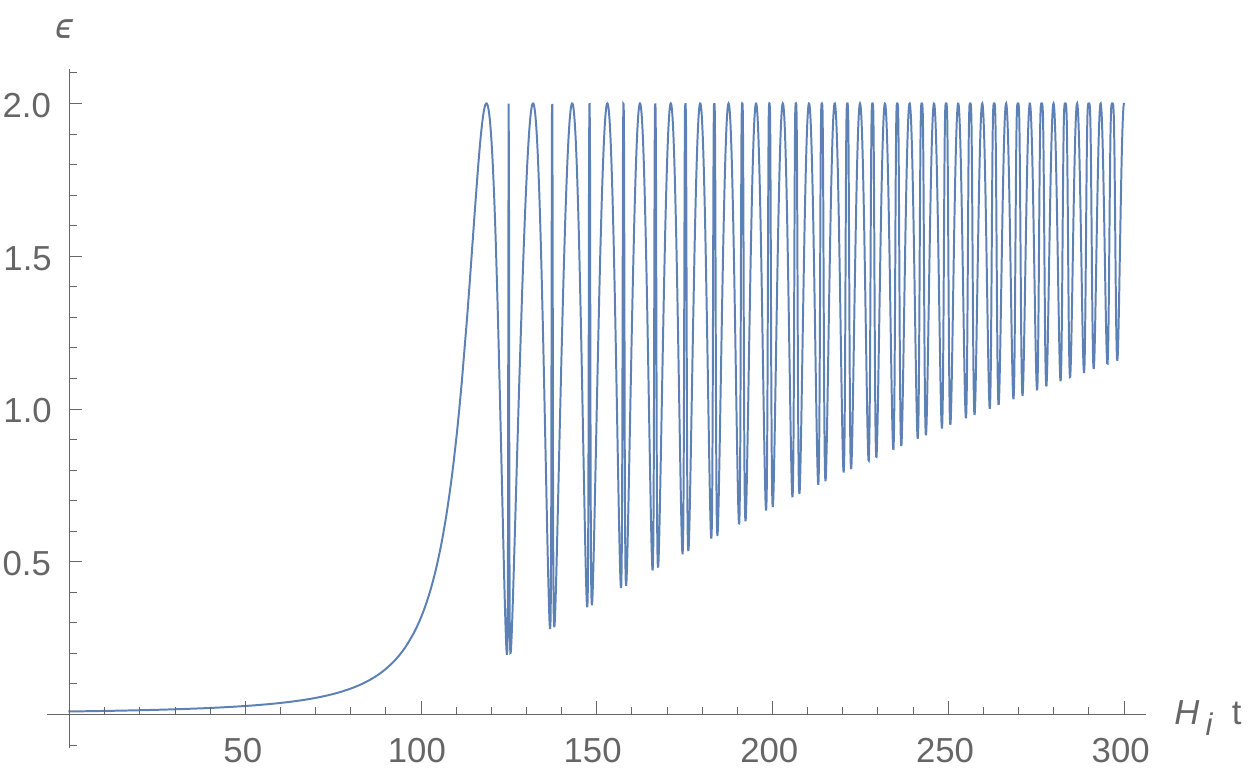}
\includegraphics[width=2.5in]{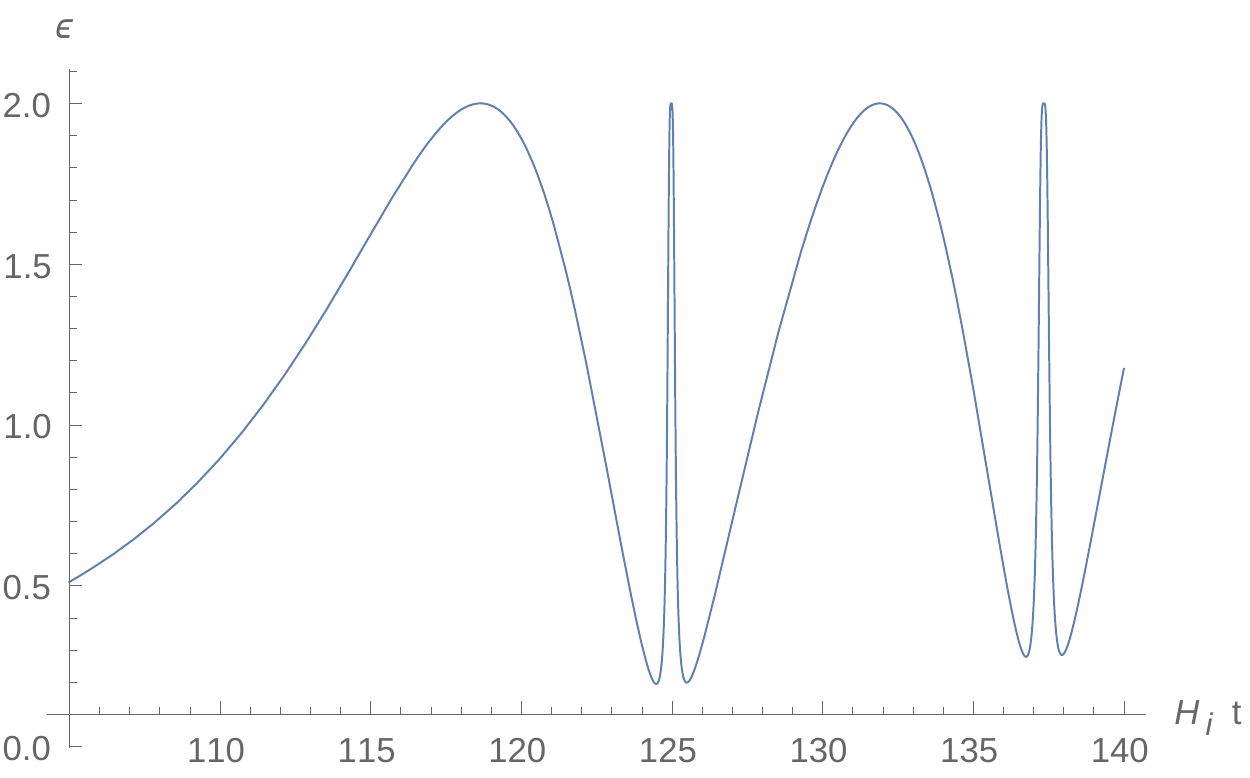}}
\centerline{\includegraphics[width=2.5in]{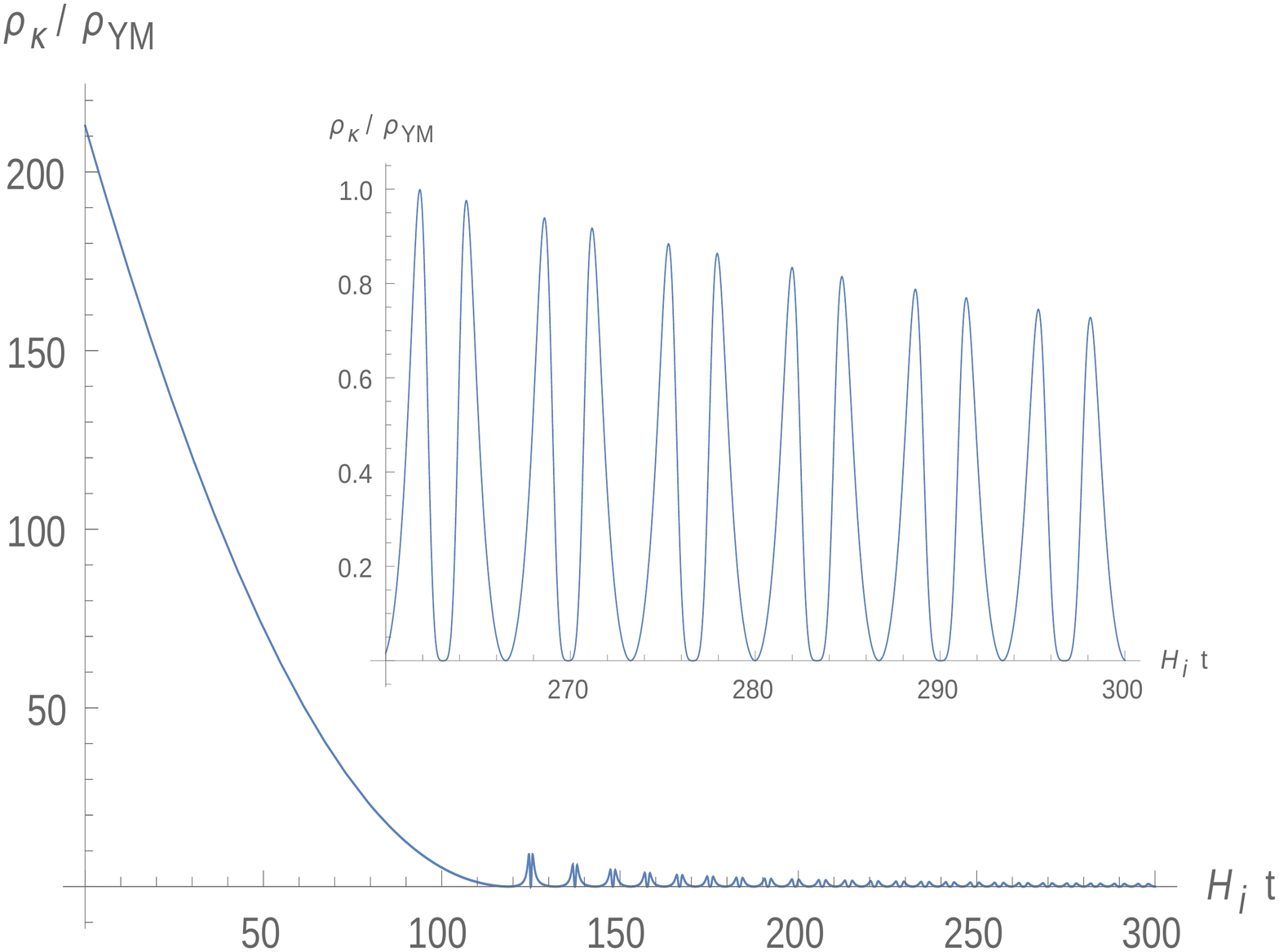}
\includegraphics[width=2.5in]{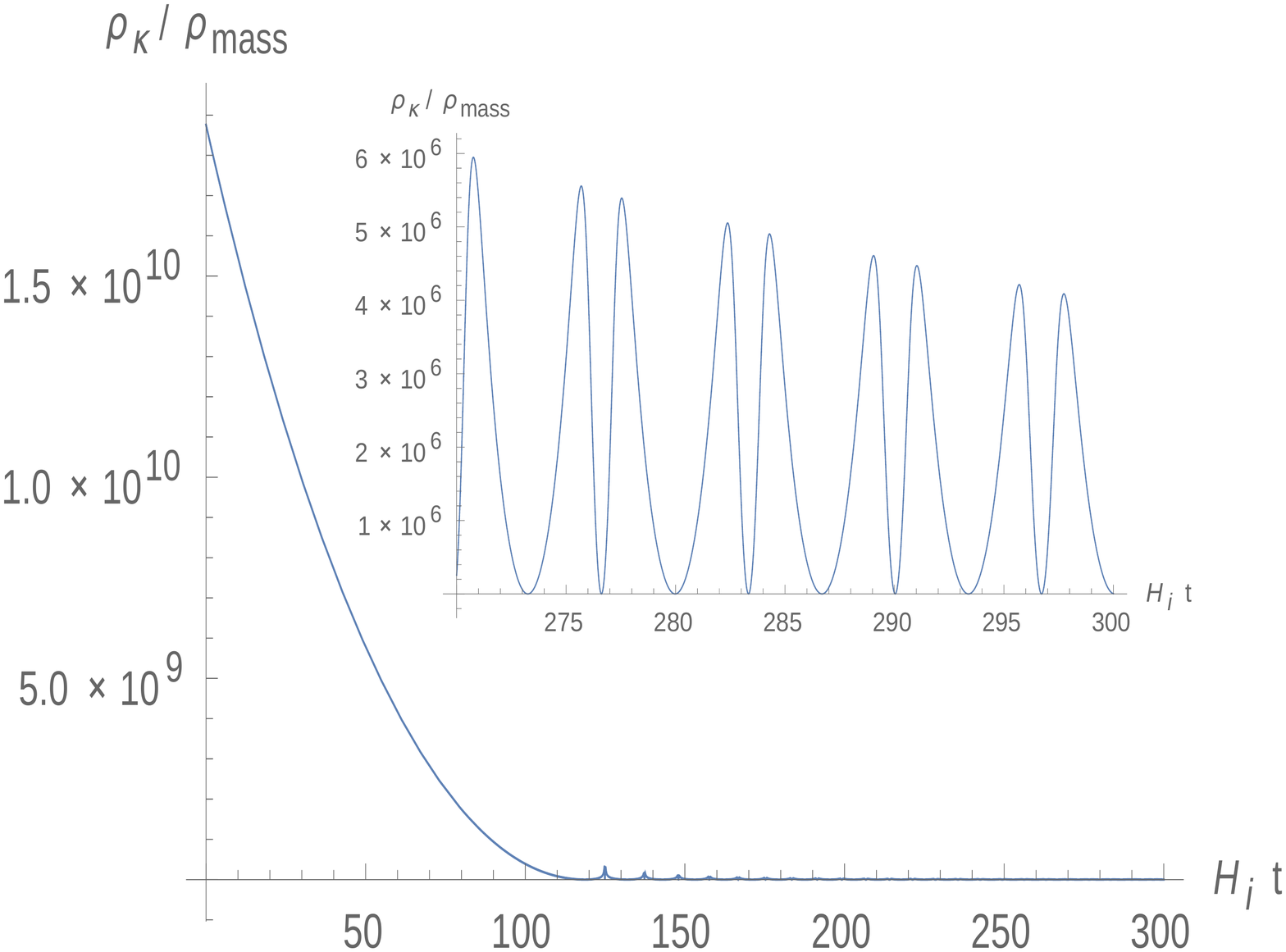}}
\centerline{\includegraphics[width=2.5in]{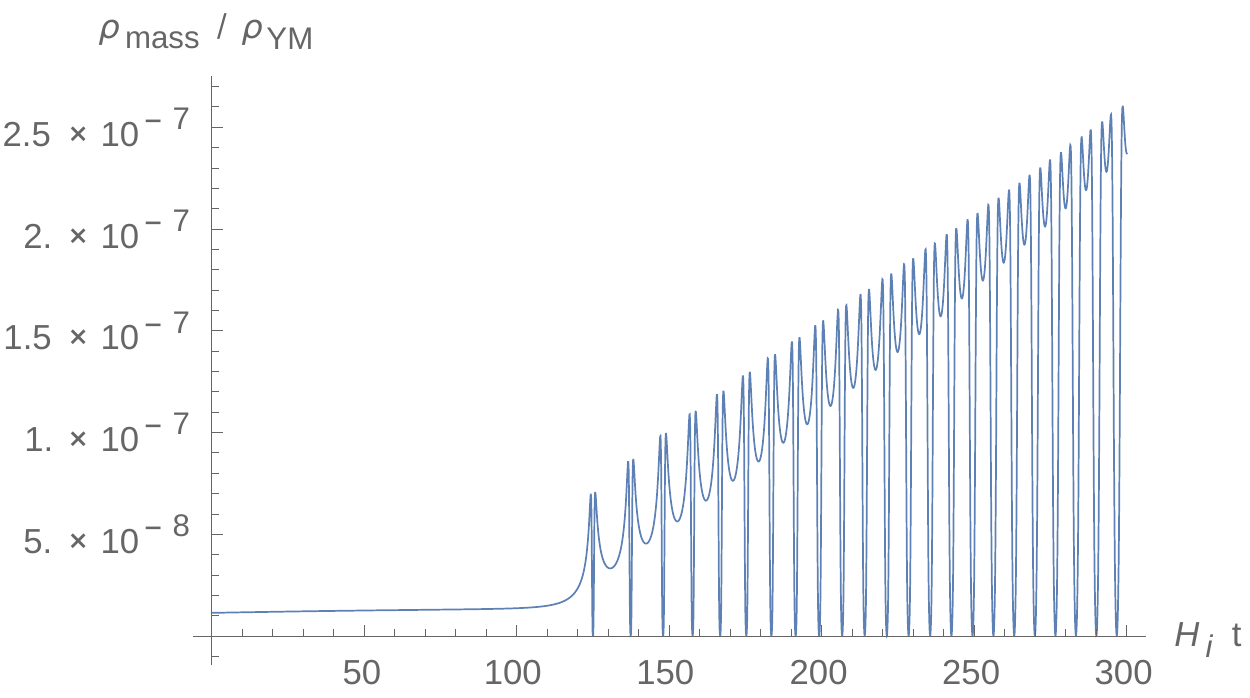}
\includegraphics[width=2.5in]{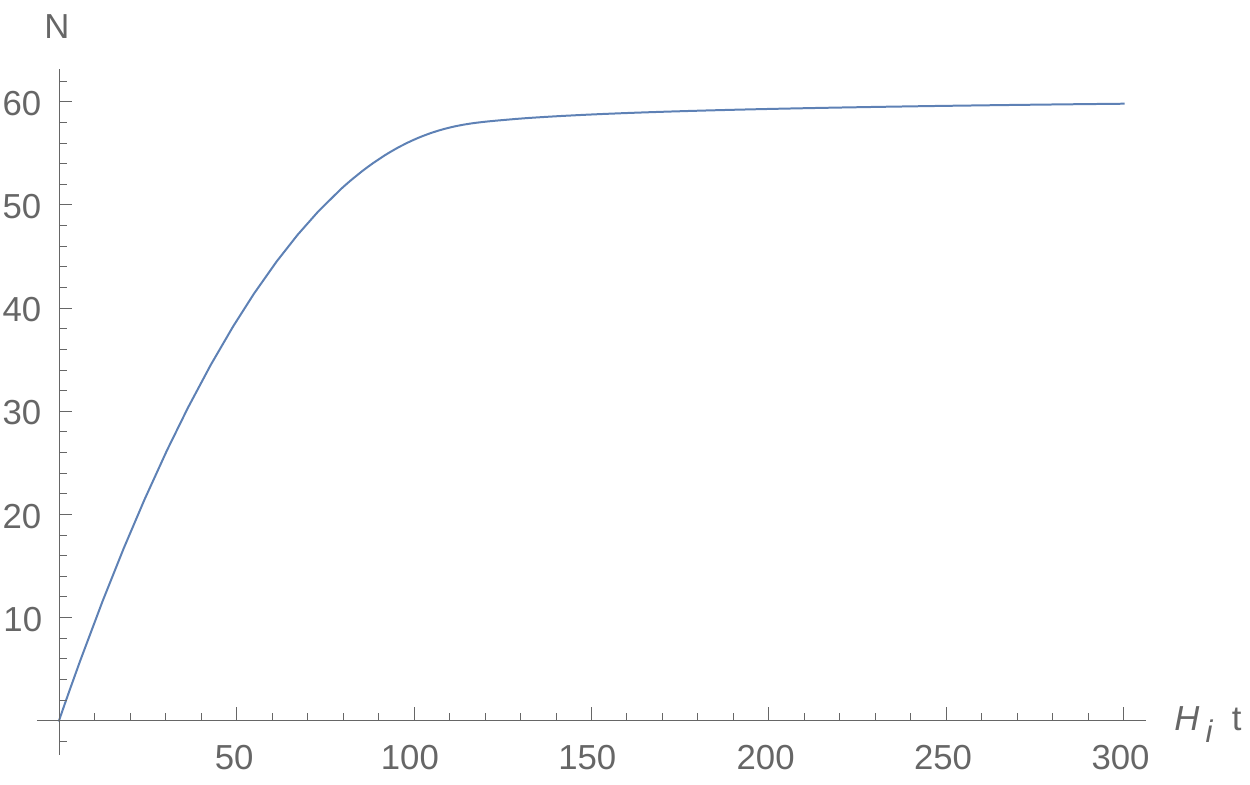}}
\vspace*{8pt}
\caption{Time evolution of $\psi$, $\varepsilon$, $\rho_{\kappa}/\rho_{YM}$, 
$\rho_{\kappa}/\rho_{mass}$,  $\rho_{mass}/\rho_{YM}$, and $N$, as well as $\dot{\phi}/a$ vs. $\psi$, found with the initial conditions $\psi_{i}=0.035$; 
$\dot{\psi}=-3.27\times10^{-9}$; $\kappa=1.733\times10^{14}$; $g=2.5\times10^{-3}$; $m=1\times10^{-8}$; 
$H_{i}=3.4\times10^{-5}$. These conditions produce the parameters $\gamma_{i}=6.59$, 
$\varepsilon_{i}=9.3\times10^{-3}$ and $\delta_{i}=1\times10^{-4}$. Note that $\kappa$, $m$, $H$ and 
$\psi$ are given in $m_{pl}$ units.  The plot of $\varepsilon$ on the right is actually a zoom of the plot on the left.  Several other panels show insets with appropriate zoomed versions.}\protect\label{fig:Firstset}
\end{figure}

The solution for the first set of initial conditions is plotted
in Fig.~\ref{fig:Firstset}.
The time evolution of $\psi$ shows slow variation at 
the beginning of inflation. This behaviour is consistent with $\delta\ll1 ,$ and then we 
infer slow-roll behaviour. This conclusion is also supported by the plot
of $\varepsilon$ showing a very small value at the beginning that 
varies slowly until inflation ends. 
Moreover, we note that the conditions for slow-roll inflation 
are satisfied since $\rho_{\kappa}\gg\rho_{YM}$ and $\rho_{\kappa}\gg\rho_{mass}$ while the
field is almost constant. 
These relations break down when the field starts to move fast and to 
oscillate.  As shown in the plot  $\dot{\phi}/a$ vs. $\psi$,   these two quantities oscillate with the same frequency but out of phase by $\pi/2$.  Nevertheless, since  the quantities $\varepsilon$, $\rho_{YM}$,  $\rho_\kappa$, and $\rho_{mass}$ involve powers of $\dot{\phi}/a$  and $\psi$, as seen in eqs. (\ref{Einstein1}), (\ref{Einstein2}), (\ref{densities}), and (\ref{Setparameters}), their behaviour is oscillatory with a modulated amplitude.  This can be seen in the the plot of $\varepsilon$ on the right (which is an appropriate zoomed version of the plot of $\varepsilon$ on the left) as well as in the zoomed plots for $\rho_\kappa/\rho_{YM}$, $\rho_\kappa/\rho_{mass}$, and $\rho_{mass}/\rho_{YM}$.  Although not included in the figure, we have numerically checked that $\dot{H}$ exhibits oscillations modulated in amplitude, whereas $H$ does not as one normally expects.
Overall, the evolution of $\psi$, $\varepsilon$ and the components of $\rho$ suggests that 
we have obtained the expected slow-roll inflationary behaviour with a sufficient
length to solve the standard 
cosmological classic problems. 
In fact, we also exhibit the evolution of the amount of inflation where 
we see that $N_{e}$ reaches a value around $60.$ Moreover, the main contribution to $N$ comes from
the slow-roll period. This result is consistent with the value from eq.~(\ref{Finalefolds}) once we 
use the initial conditions to obtain 
$\gamma_{i}$, $\varepsilon_{i},$ and $\omega_{i}$.  We assumed a small value for $m$ compared to $H_{i}$ (i.e., $\omega_{i}$)  for the first set 
of initial conditions. 
Therefore, the model is not drastically affected by the mass term (the ratio  $\rho_{mass}/\rho_{YM}$ always stays negligible despite its pronounced oscillations). We obtain almost the same behaviour as
found for the original model. Nevertheless, it is also possible to use a value of $m$ larger 
than $H_{i}$ and obtain inflation as shown below.

\begin{figure}[h]
\centerline{\includegraphics[width=2.5in]{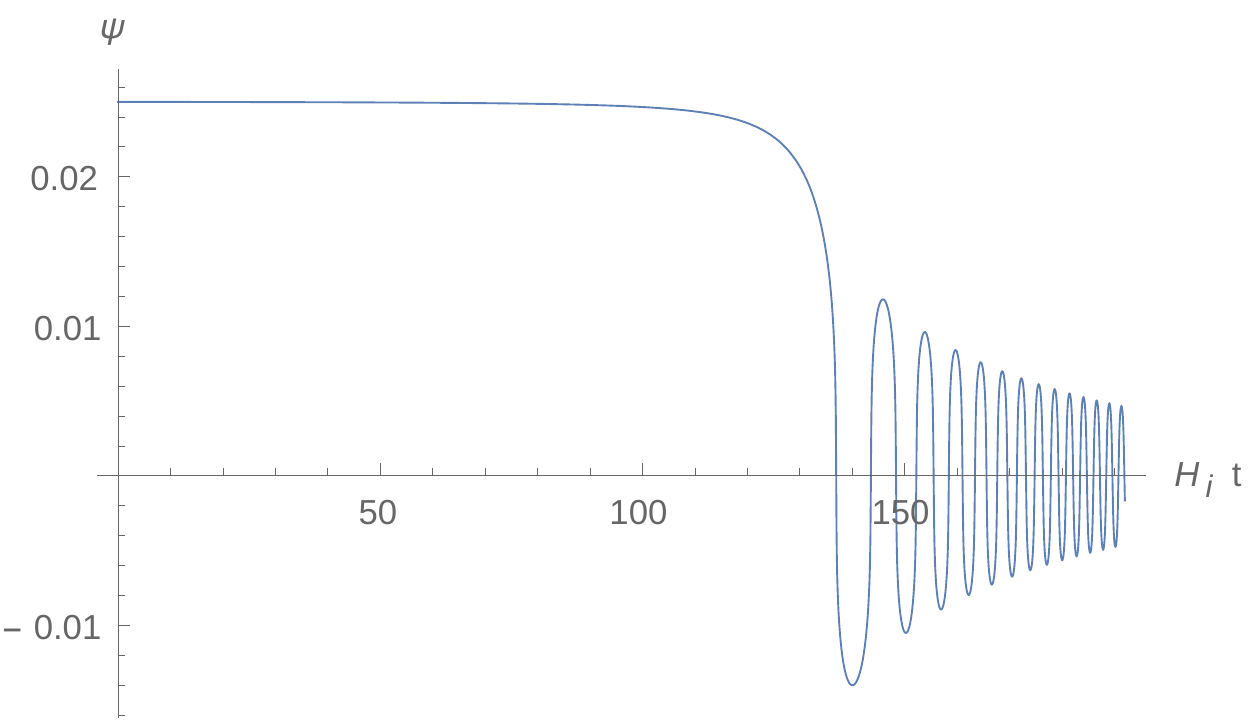}
\includegraphics[width=2.5in]{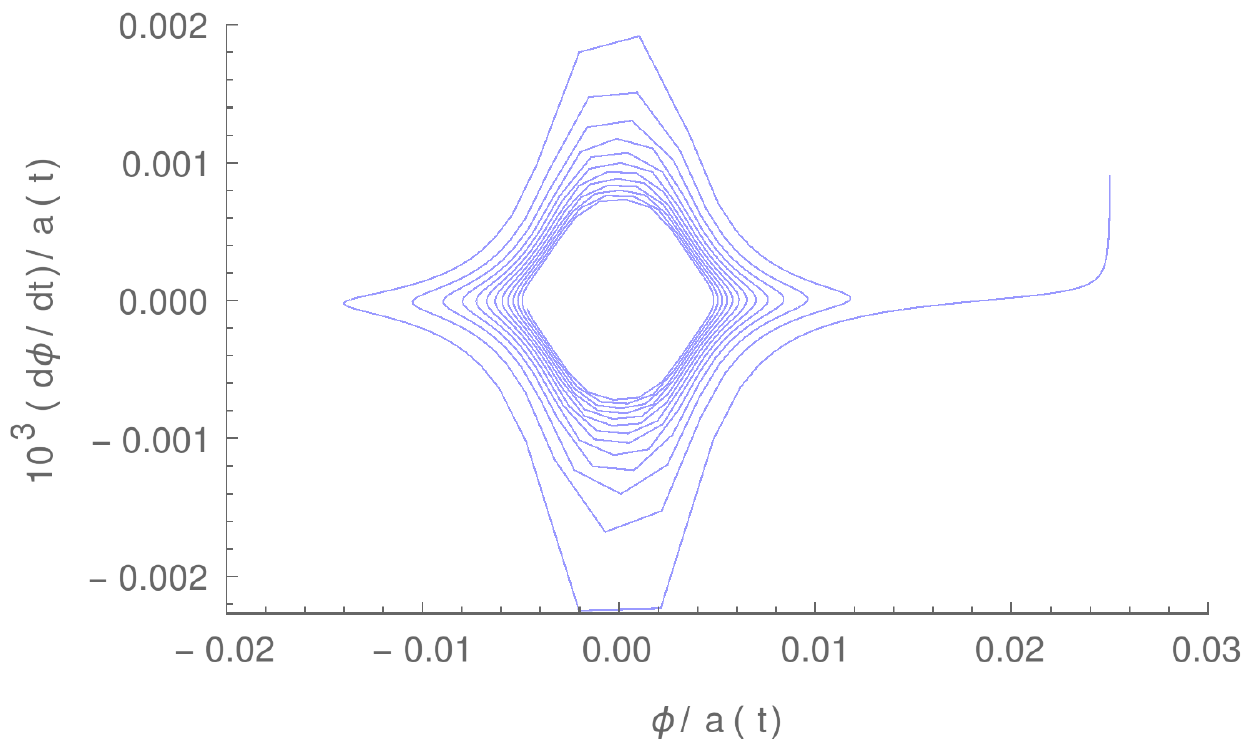}} 
\centerline{\includegraphics[width=2.5in]{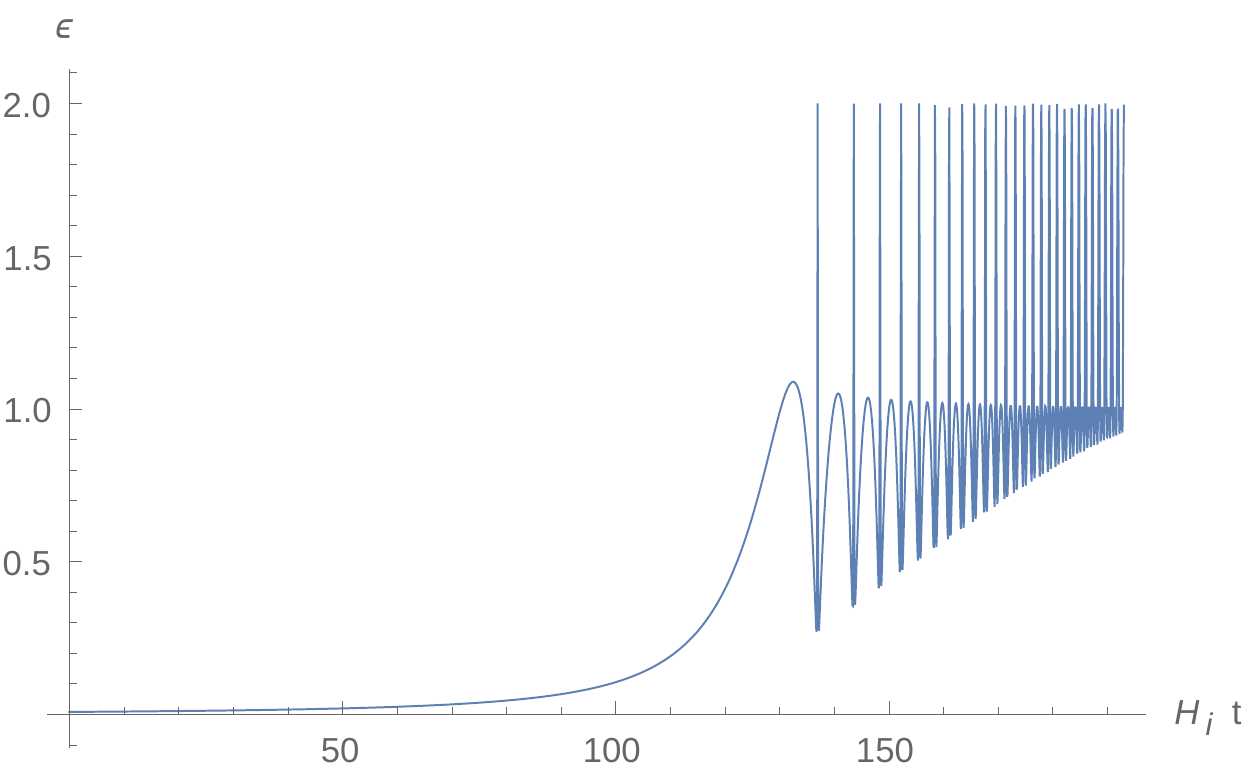}
\includegraphics[width=2.5in]{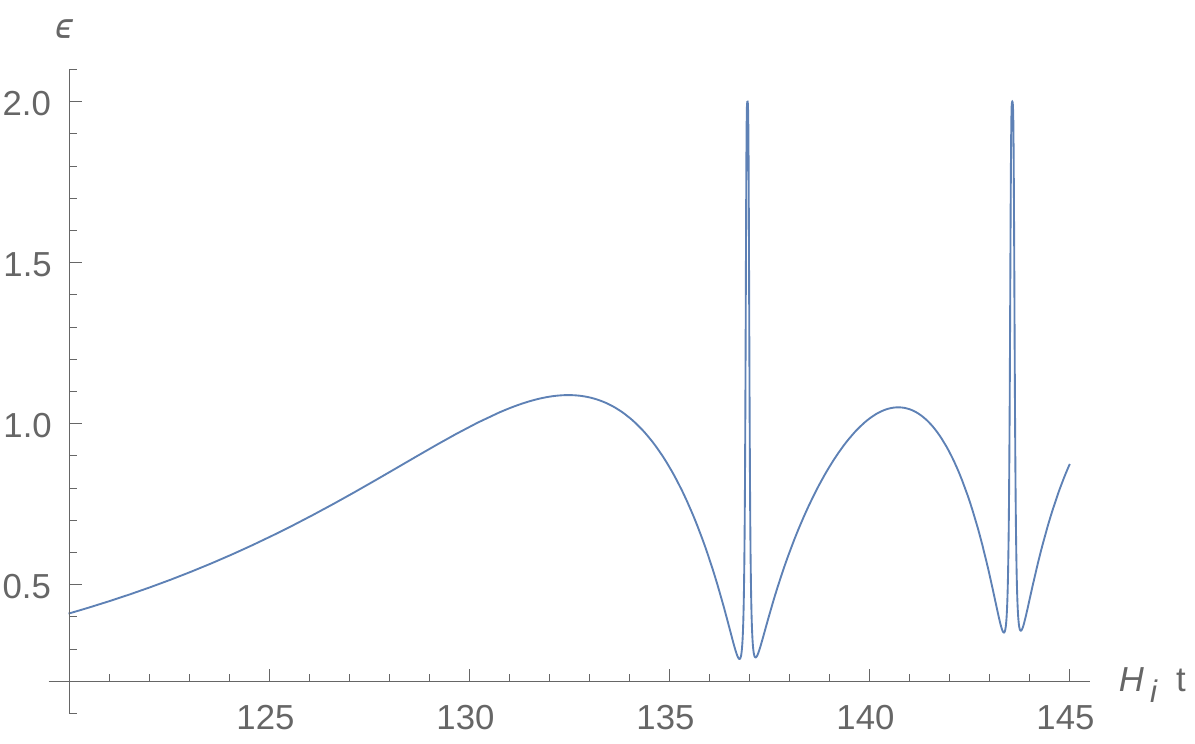}}
\centerline{\includegraphics[width=2.5in]{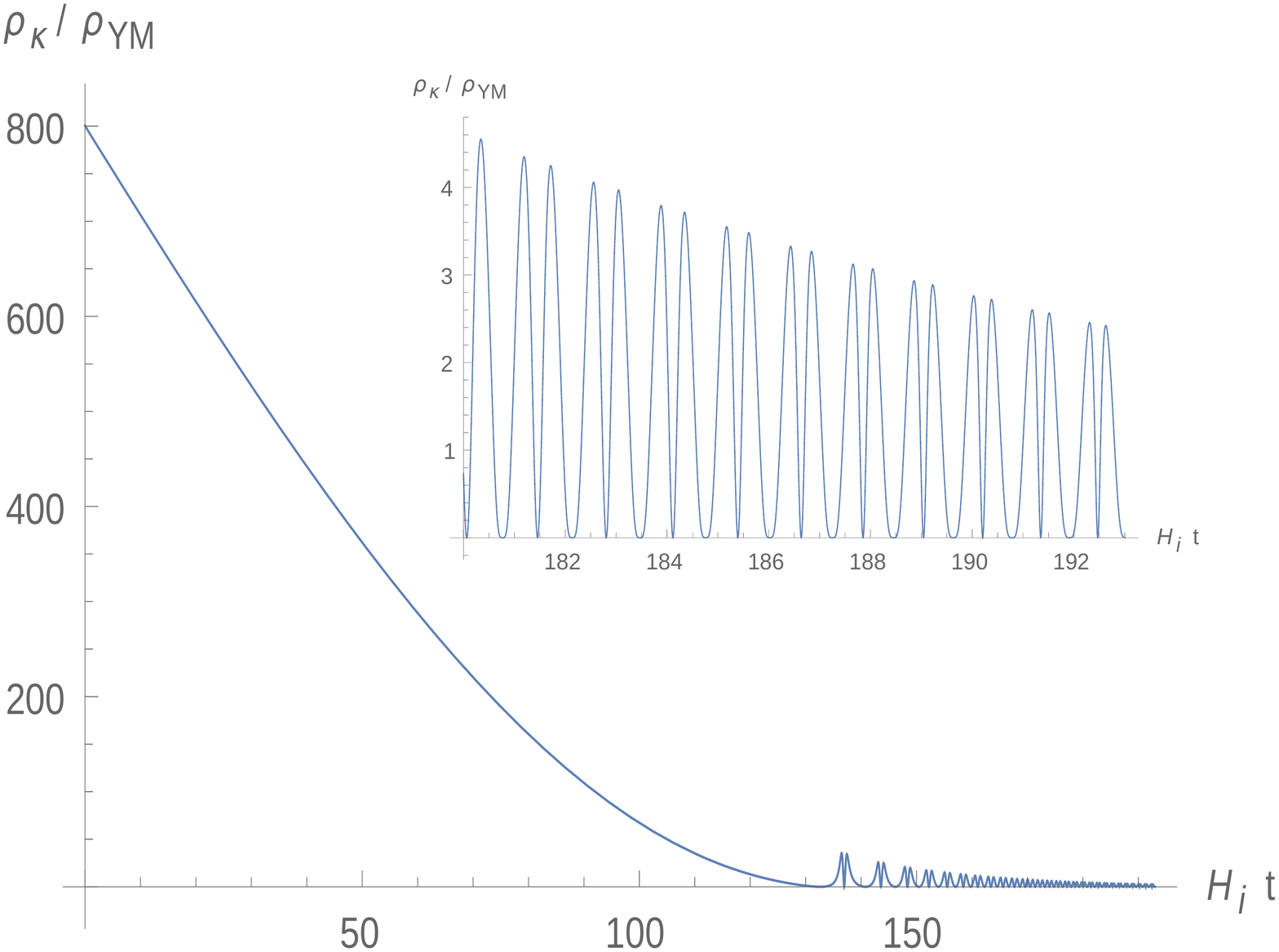}
\includegraphics[width=2.5in]{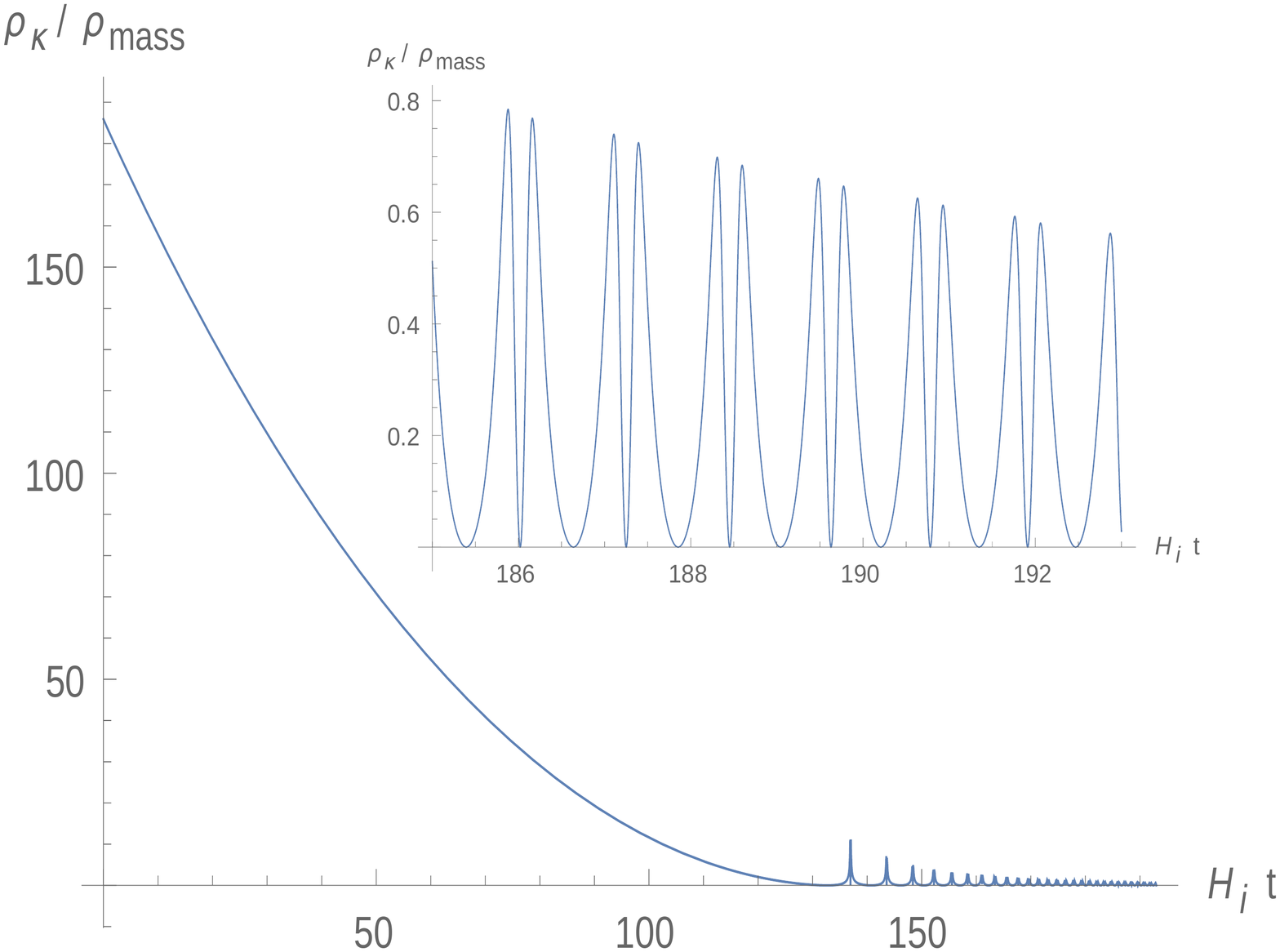}}
\centerline{\includegraphics[width=2.5in]{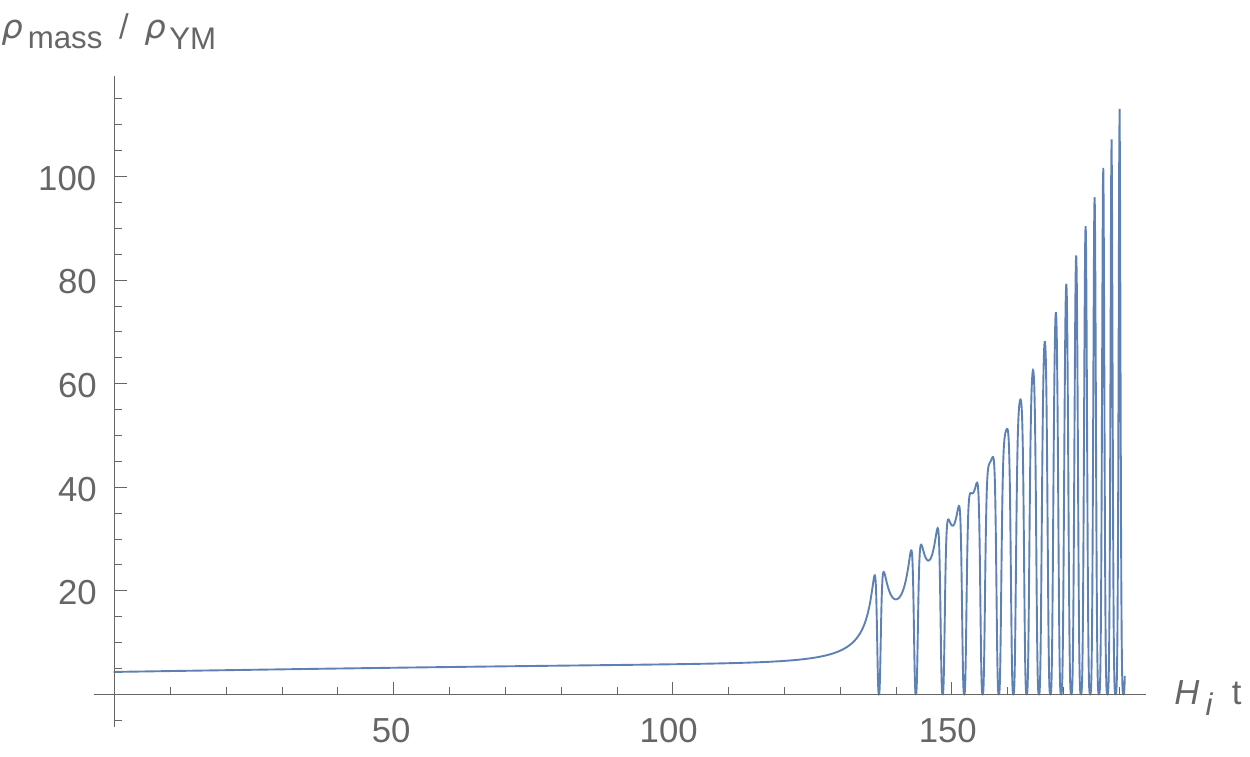}
\includegraphics[width=2.5in]{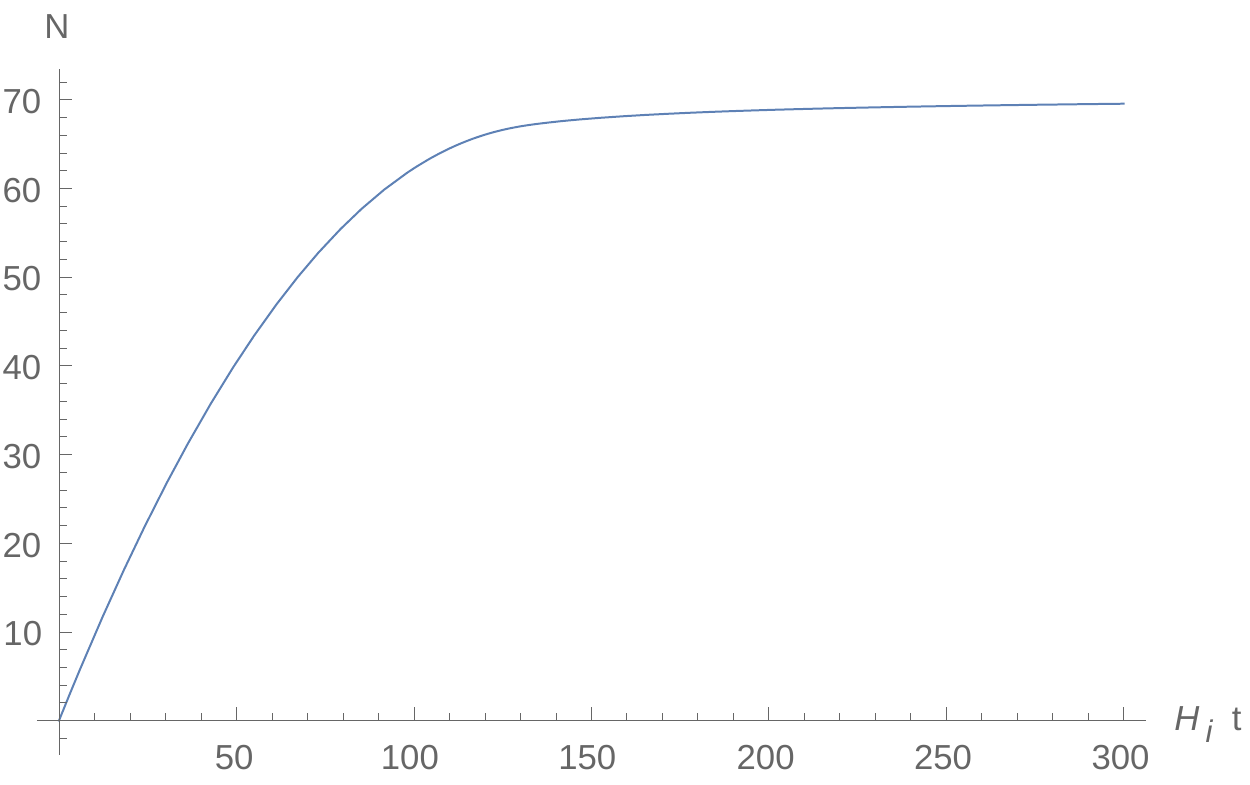}}
\vspace*{8pt}
\caption{Time evolution of $\psi$, $\varepsilon$, $\rho_{\kappa}/\rho_{YM}$, 
$\rho_{\kappa}/\rho_{mass}$,  $\rho_{mass}/\rho_{YM}$, and $N$, as well as $\dot{\phi}/a$ vs. $\psi$, found with the initial conditions $\psi_{i}=0.025$; 
$\dot{\psi}=9.07\times10^{-9}$; $\kappa=1.3\times10^{15}$; $g=2.5\times10^{-3}$; $m=1.5\times10^{-4}$; 
$H_{i}=3.6\times10^{-5}$. These conditions produce the parameters $\gamma_{i}=2.96$, 
$\varepsilon_{i}=7.8\times10^{-3}$ and $\delta_{i}=-8\times10^{-4}$. Note that $\kappa$, $m$, $H$ and 
$\psi$ are given in $m_{pl}$ units. The plot of $\varepsilon$ on the right is actually a zoom of the plot on the left.  Several other panels show insets with appropriate zoomed versions.\protect\label{fig:Secondset}}
\end{figure}

For the second set of initial conditions and parameters, we take $m>H_{i}$ (see Fig.~\ref{fig:Secondset}). 
In this case, all the quantities evolve with essentially the same shape 
as in Fig.~\ref{fig:Firstset}, except for $\varepsilon$ and $\rho_{mass}/\rho_{YM}$. During an initial period, the field $\psi$ varies slowly 
generating an amount of inflation above the lower bound, taking into account that we have to increase 
$\kappa$ in order to have a sufficiently long $\rho_{\kappa}$ dominance. The initial conditions and 
parameters are suitable for describing a slow-roll inflationary period. As in the previous case, 
we checked the validity of eq.~(\ref{Finalefolds}) for calculating the final amount of inflation.
We found the value $67,$ which is in agreement with the plot in Fig.~\ref{fig:Secondset}.
As in the previous figure, the quantities $\varepsilon$, $\rho_{YM}$,  $\rho_\kappa$, and $\rho_{mass}$ exhibit oscillations modulated in amplitude. The curious behaviour of $\varepsilon$, with alternating peaks of oscillation, can be explained by the high mass value:  the plot for $\rho_{mass}/\rho_{YM}$ reveals a series of oscillations where the energy density is alternately dominated by $\rho_{mass}$, giving $\varepsilon$ values quite near 1, and by  $\rho_{YM}$, giving $\varepsilon$ a value equal to 2.   More specifically, since $\rho_{mass}$ and $\rho_\kappa$ vanish periodically while $\rho_{YM}$ never vanishes, according to eq. (\ref{densities}), because of the $\pi/2$ phase difference between the oscillations of $\phi$ and $\dot{\phi}$, the highest peaks in $\varepsilon$ always reach 2 according to eq. (\ref{Condslow}) ($\phi = 0$ so that $\rho_{mass} = \rho_\kappa = 0$) whereas the lowest peaks reach values near 1 ($\rho_{mass} \gg \rho_{YM} \gg \rho_\kappa$).

\begin{figure}[H]
\centerline{\includegraphics[width=2.5in]{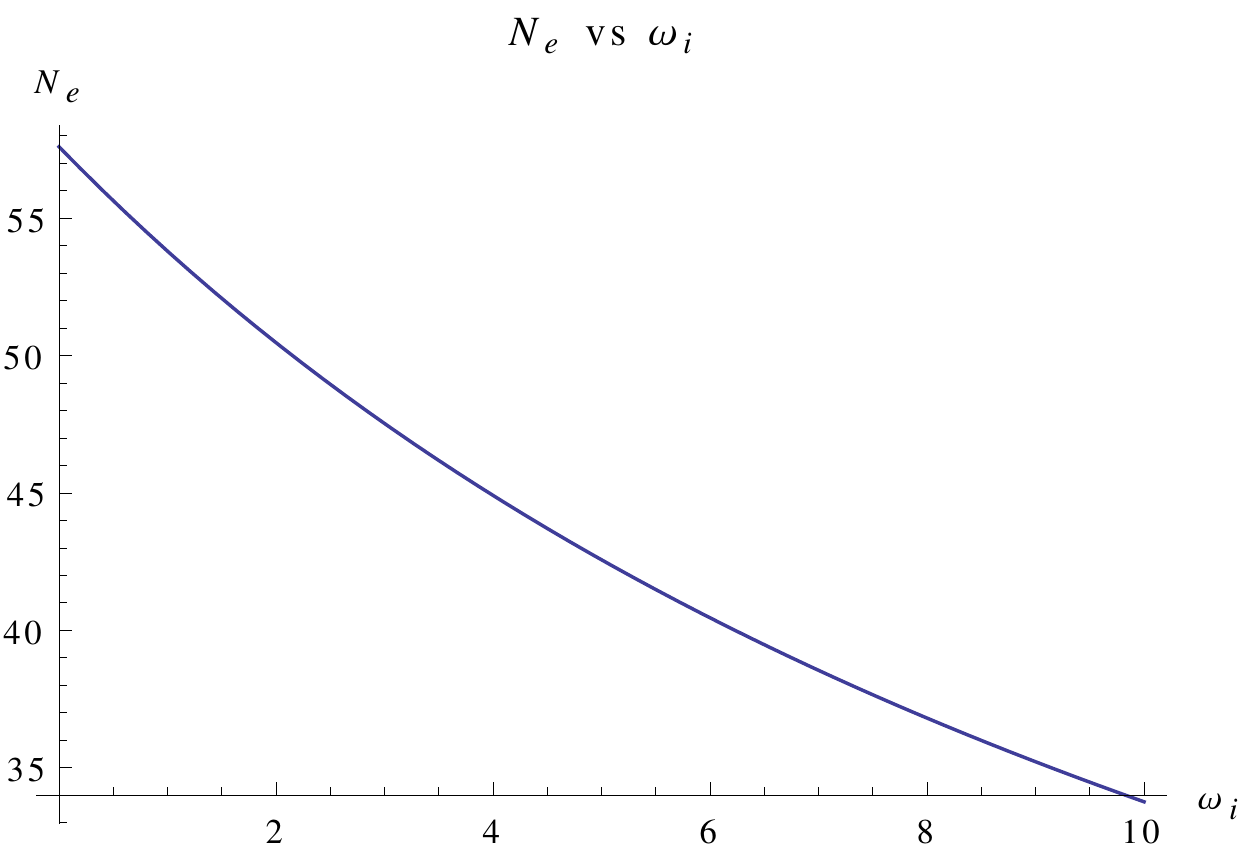}
\includegraphics[width=2.5in]{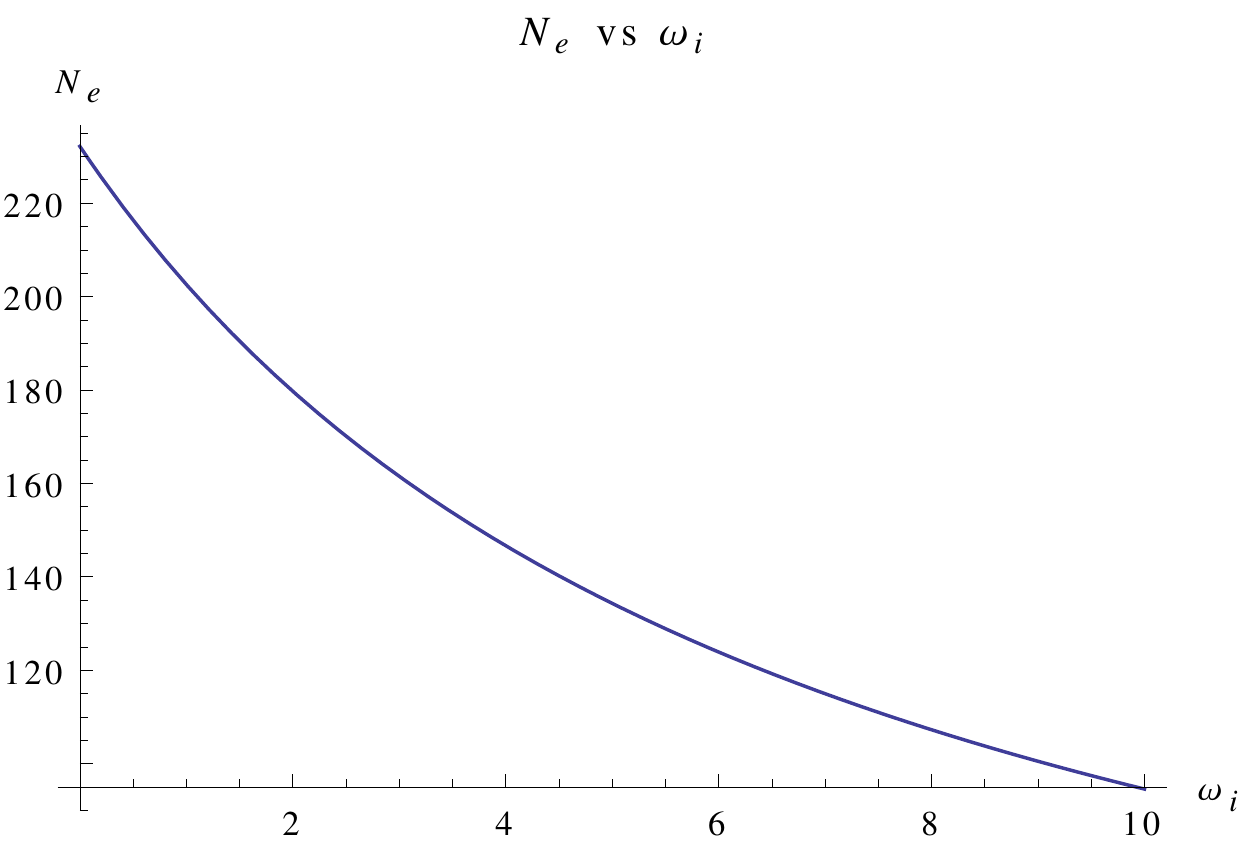}
}
\vspace*{8pt}
\caption{Change in the final amount of inflation $N_{e}$, given by eq. (\ref{Finalefolds}), as a 
function of $\omega_{i}$ for the two sets of initial conditions used in Figs. \ref{fig:Firstset} and \ref{fig:Secondset} (left and right respectively).\protect\label{fig:Nwan}}
\end{figure}

Finally, by plotting $N_{e}$ as a function of 
$\omega_{i}=m^{2}/H_{i}^{2}$ (keeping $H_{i}$ constant) for the two sets of initial conditions 
(see Fig.~\ref{fig:Nwan}), we find that $N_{e}$ decreases substantially
provided that $m$ increases above 
$H_i$.  This shows how the mass term modifies the original massless Gauge-flation model: it reduces the 
length of inflation.
 
\section{Conclusions}\label{sec:con}

Including the mass term in the Gauge-flation model presents interesting features analysed 
throughout the paper and allows us to obtain a period of accelerated expansion. We have found that the 
presence of inflationary expansion does not depend on the mass of the fields. The mathematical 
condition for inflation is $\rho_{\kappa}>\rho_{YM}$ in which the mass contribution does
not appear.
However, slow-roll inflation does not occur unless 
$\rho_{\kappa}\gg\rho_{mass}$, i.e., the mass term contribution to the energy 
density $\rho$ must be negligible compared to the $\kappa$-term contribution at the beginning of inflation. 
This is because the $\kappa$-term produces the equation of state $P=-\rho$ 
leading to an era of accelerated expansion. By analysing the numerical solution, we observe that 
the conditions $\rho_{\kappa}\gg\rho_{YM}$ and $\rho_{\kappa}\gg\rho_{mass}$ hold for some initial time 
after which these inequalities are inverted, causing 
inflation comes to an end (when $\varepsilon\rightarrow1$). We found that it is possible to choose
sets of initial conditions and parameters of the model so 
that inflation lasts long enough to solve the classic problems of the standard cosmology. By using the 
slow-roll approximation, we found an analytic expression for the amount of inflation in 
terms of $\gamma_{i}$, $\varepsilon_{i}$ and $\omega_{i}$. The validity of this expression was checked 
using a numerical solution, which was useful for clarifying the influence
of the mass term on the 
model dynamics. 
Fig.~\ref{fig:Nwan} shows that an increase in $m$ results in a smaller $N_{e}$ (i.e., the 
mass term is relevant at determining the length of inflation). We conclude that the new model 
provides a successful setup for the description of the primordial inflationary evolution of the 
Universe. 
A study of the first-order perturbations will be reported in a future paper 
illustrating the impact of the longitudinal mode of the field perturbations on the graph $r$ vs. $n_s$ 
and on the tachyonic instability at $\gamma < 2$.  We expect an enhancement in the  $\eta$ parameter because of the mass term as happens in single-field inflation, and, consequently, an enhancement in $n_s$.  Thus, when $\gamma$ is small, but still in the stable region (and we will have to check how it changes when the mass is introduced), $n_s$ would not continue being too low and, therefore, the pair $r$ - $n_s$ would lie in the allowed parameter window.

\section*{Acknowledgments}

We are indebted to Martin Bucher and Fabio Lora for interesting criticisms and helpful advice.  This work was supported by COLCIENCIAS - ECOS NORD grant number RC 0899-2012 with the help of ICETEX, 
and by COLCIENCIAS grant numbers 110656933958 RC 0384-2013 and 123365843539 RC FP44842-081-2014.

\end{document}